\documentstyle[preprint,aps]{revtex}
\begin{document}
\draft
\title{A comparative study of super- and highly-deformed bands 
in the $A\sim 60$ mass region}
\author{A.\ V.\ Afanasjev$^{1,2,3}$, I.\ Ragnarsson$^{3}$, P.\
Ring$^{1}$}
\address{$^{1}$ Physik-Department der Technischen Universit{\"a}t 
M{\"u}nchen,  D-85747 Garching, Germany,\\
$^2$Nuclear Research Center, Latvian Academy of Sciences,
LV-2169, Salaspils, Miera str. 31, Latvia, \\
$^3$Department of Mathematical Physics, 
Lund Institute of Technology, 
PO Box 118, S-22100, Lund, Sweden}
\maketitle
\date{\today}
\maketitle
\begin{abstract}
Super- and highly-deformed rotational bands
in the $A\sim 60$ mass region are studied within 
cranked relativistic mean field theory and the 
configuration-dependent shell-correction approach based on 
the cranked Nilsson potential. Both approaches describe 
the experimental data well. Low values of the dynamic 
moments of inertia $J^{(2)}$ compared with the kinematic 
moments of inertia $J^{(1)}$ seen both in experiment and 
in calculations at high rotational frequencies indicate 
the high energy cost to build the states at high spin and 
reflect the limited angular momentum content in these
configurations.
\end{abstract}
\vspace{2.0cm}
\input epsf

\baselineskip=20.9pt
Extremely fast rotating nuclei are interesting 
laboratories providing information for test of theoretical 
models at extreme conditions (large angular momentum 
and/or deformation, limit of angular momentum in
the rotational bands etc.). A special feature at the
high rotational frequencies is that pairing correlations
are considerably quenched and can often be neglected.
A most interesting nuclear region is the one with 
$A\sim 60\,\, (N\approx Z \approx 30)$, 
where a large variety of rotational 
structures such as (smooth) terminating bands, 
highly-deformed and superdeformed (SD) rotational bands are 
expected to be observed up to very high rotational frequencies 
in the same nucleus. The $^{62}$Zn nucleus \cite{Zn62SD,Zn62bt} 
represents a first example of this variety.

Of special interest are the properties of SD bands in this region
since they extend to the highest rotational frequencies ($\sim 1.8$ MeV) 
observed so far in SD bands. The fact that the predicted 
SD band in the doubly-magic superdeformed nucleus $^{60}$Zn 
\cite{Rag90} has been observed \cite{Zn60SD} and that it is 
linked to the low-spin level scheme is another attractive point. 
This is because by means of an effective alignment (or similar) 
approach \cite{Rag91,Rag93}, it becomes possible to map not only 
relative spin values as in the $A\sim 140-150$ mass region 
(see Refs.\ \cite{Rag93,ALR.98}) but also absolute spin values in the 
unlinked SD and highly-deformed bands. Then, in SD bands with 
little influence of pairing correlations, it will be possible 
to make a comparison between experiment and theory, 
not only for those physical observables which can be extracted 
without knowing absolute spin values (like dynamic moment of inertia 
$J^{(2)}$ and the transition quadrupole moment $Q_t$)
but also, for the first time, for observables which cannot be 
extracted without such a knowledge (like the kinematic moment 
of inertia $J^{(1)}$ and the evolution of the excitation 
energy within a band as a function of spin).

In the present article, a comparative study of the recently 
observed highly-deformed band in $^{58}$Cu \cite{Cu58} and
the SD bands in $^{60,62}$Zn \cite{Zn62SD,Zn60SD} is 
presented. In addition, the general features of SD and 
highly-deformed bands in this mass region of $A\sim 60$ are 
outlined. Our theoretical tools are the cranked relativistic 
mean field theory (further CRMF) \cite{KR.89,KR.90} and the 
configuration-dependent shell-correction approach based on the 
cranked Nilsson potential (further CN) \cite{Beng85,A110}. 

In relativistic mean field (RMF) theory the nucleus is described 
as a system of point-like nucleons represented by Dirac-spinors
and coupled to mesons and to the photon. The nucleons interact 
by the exchange of several mesons, namely, the scalar $\sigma$ 
and three vector particles $\omega$, $\rho$ and the photon. 
The CRMF theory represents the extension of relativistic mean 
field (RMF) theory to the rotating frame.  CRMF theory is a 
fully self-consistent theory. On the contrary, in the CN 
model the total energy is described as a sum of the rotating 
liquid drop energy and the shell correction energy. This 
leaves some room for inconsistencies between 
macroscopic and microscopic parts as illustrated for 
example in Refs.\ \cite{KRA.98,Dud84}. However, it is 
commonly accepted that the CN model provides a reasonable 
description of the nuclear many-body problem. Both 
models have been very successful in describing different 
aspects of SD bands in the $A\sim 140-150$ mass 
region (see e.g. \cite{ALR.98,AKR.96} and \cite{Rag93,KRA.98}).
The details of the formalisms of these two approaches can 
be found in Refs.\ \cite{ALR.98,AKR.96} and in 
Refs.\ \cite{Beng85,A110}, respectively. 

CRMF calculations have been performed with three parameterizations 
of the RMF Lagrangian (NL1 \cite{NL1},
NL3 \cite{NL3} and NLSH \cite{NLSH}) in order to define the 
force best suited for the description of rotational properties 
of the nuclei with $N\approx Z$. Since the results with NL3 
are rather close to the ones with NLSH, they are not shown 
in some figures. The spatial components of the vector 
mesons {(\it nuclear magnetism}) play an extremely 
important role for the description of moments of inertia 
\cite{KR.93}. They are taken into account in a fully self-consistent 
way. 

The relativistic mean field equations are solved in the basis
of a deformed harmonic oscillator. A basis deformation of 
$\beta_0=0.2$ has been used. All bosonic states below the energy
cutoff $E^{cut-off}_B \leq 16.5\hbar\omega^B_0$ and all
fermionic states below the energy cutoff $E^{cut-off}_F 
\leq 13.5\hbar\omega^F_0$ have been used in the diagonalization. 
The increase of the fermionic space compared with the truncation 
scheme used in Ref.\ \cite{ALR.98} was necessary because in the 
present study, we compare experimental and calculated excitation 
energies relative to rigid rotor reference which requires high 
accuracy in the calculation of energies. Note, however, that the 
energy cutoff $E^{cut-off}_F \leq 11.5\hbar\omega^F_0$ provides 
a rather good description of moments of inertia and the 
quadrupole and hexadecupole moments and thus it can be 
used for a more systematic investigations. 

  In the CN calculations, the standard set of the parameters for 
the Nilsson potential \cite{Beng85} has been used. In both approaches, 
pairing correlations are not taken into account. Therefore, the 
results can be considered as realistic only in the region of high 
spins, say $I\geq 15\hbar$. However, for some configurations the 
paired band crossings at low spin will be blocked and thus, in 
these cases, the results of the calculations are not expected to 
deviate significantly from experiment even at lower spin values; 
for details see the discussion in Ref.\ \cite{AKR.96}. To label
the configurations we use the shorthand notation $[p_1p_2,n_1n_2]$ 
where $p_1$ $(n_1)$ is the number of proton (neutron) 
$f_{7/2}$ holes and $p_2$ $(n_2)$ is the number of proton 
(neutron) $g_{9/2}$ particles. Superscripts to the orbital 
labels are used to indicate the sign of signature $r$ 
($r=\pm i$) of the orbital. 

\begin{figure}
\epsfxsize 12.5cm
\epsfbox{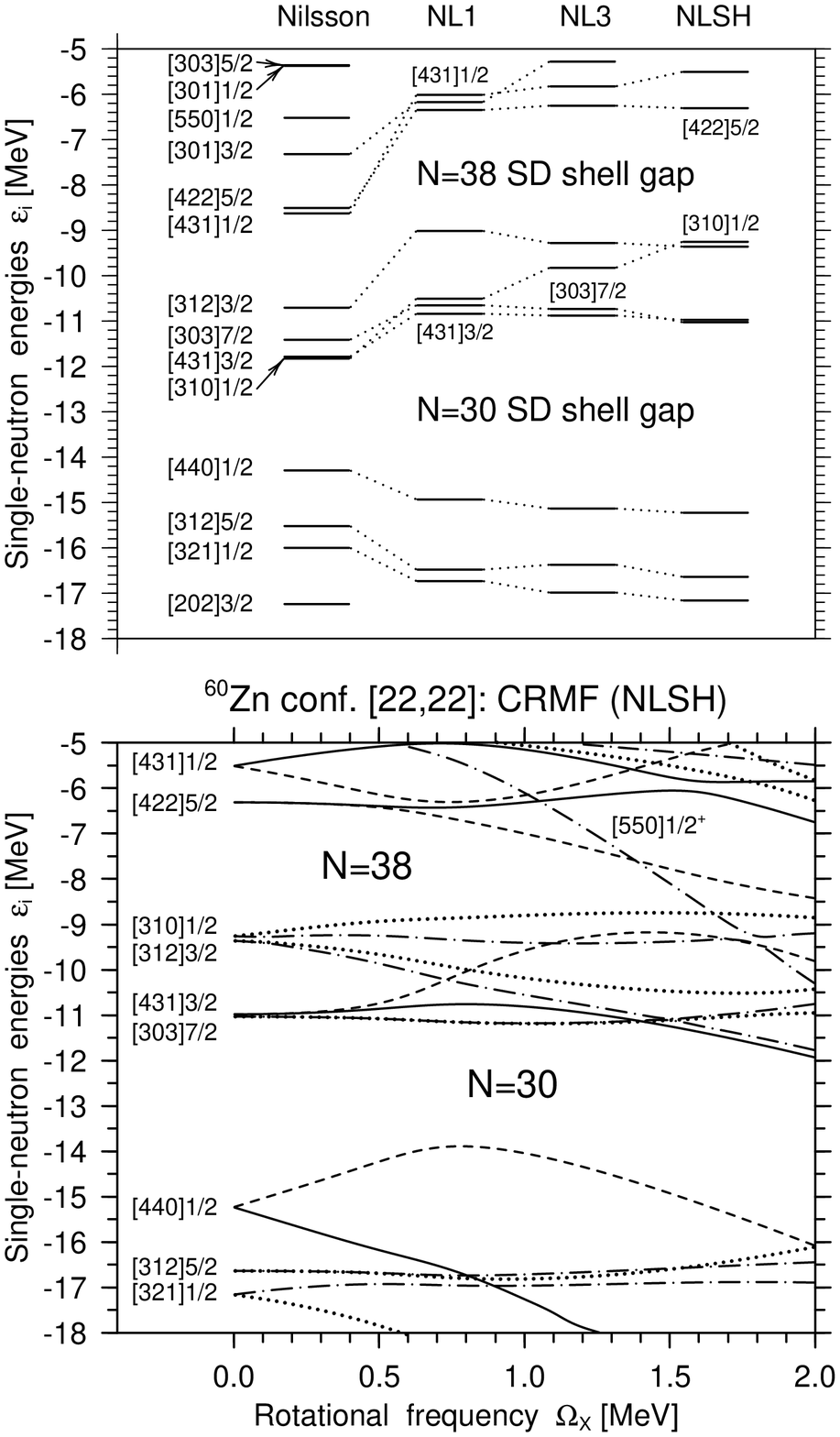}
\caption{Bottom panel: Neutron single-particle energies 
(routhians) in the self-consistent rotating potential as a 
function of the rotational frequency $\Omega_x$ calculated in 
CRMF theory with parameter set NLSH. They are given 
along the deformation path of the lowest SD configuration 
[22,22] in $^{60}$Zn. Solid, short-dashed, dot-dashed
and dotted lines indicate $(\pi=+,r=-i)$, 
$(\pi=+,r=+i)$, $(\pi=-,r=+i)$ and $(\pi=-,r=-i)$ orbitals,
respectively. At $\Omega_x=0.0$ MeV, the single-particle
orbitals are labelled by the asymptotic quantum
numbers $[Nn_z\Lambda]\Omega$ (Nilsson quantum numbers)
of the dominant component of the wave function.
Top panel:
The single-particle states around the $N=30$
SD shell gap calculated with the Nilsson potential and
three parameterizations of RMF theory at the corresponding
equilibrium deformations of the [22,22] configuration in 
$^{60}$Zn at $\Omega_x =0.0$ MeV. 
It is only in the CRMF calculations that the energies  
are absolute, in the Nilsson potential the energies are shifted 
so that the $N=30$ SD shell gaps coincide roughly in both 
approaches. The relative single-particle energies
are approximately the same for the protons as for the
neutrons, but the absolute proton energies are higher
because of the Coulomb energy. The fact that the spectrum
is less dense in RMF theory than in the Nilsson
potential is related to low effective mass 
$(m^*/m\approx 0.6)$ in RMF theory.}
\label{routh-crmf}
\end{figure}

\begin{figure}
\epsfxsize 12.5cm
\epsfbox{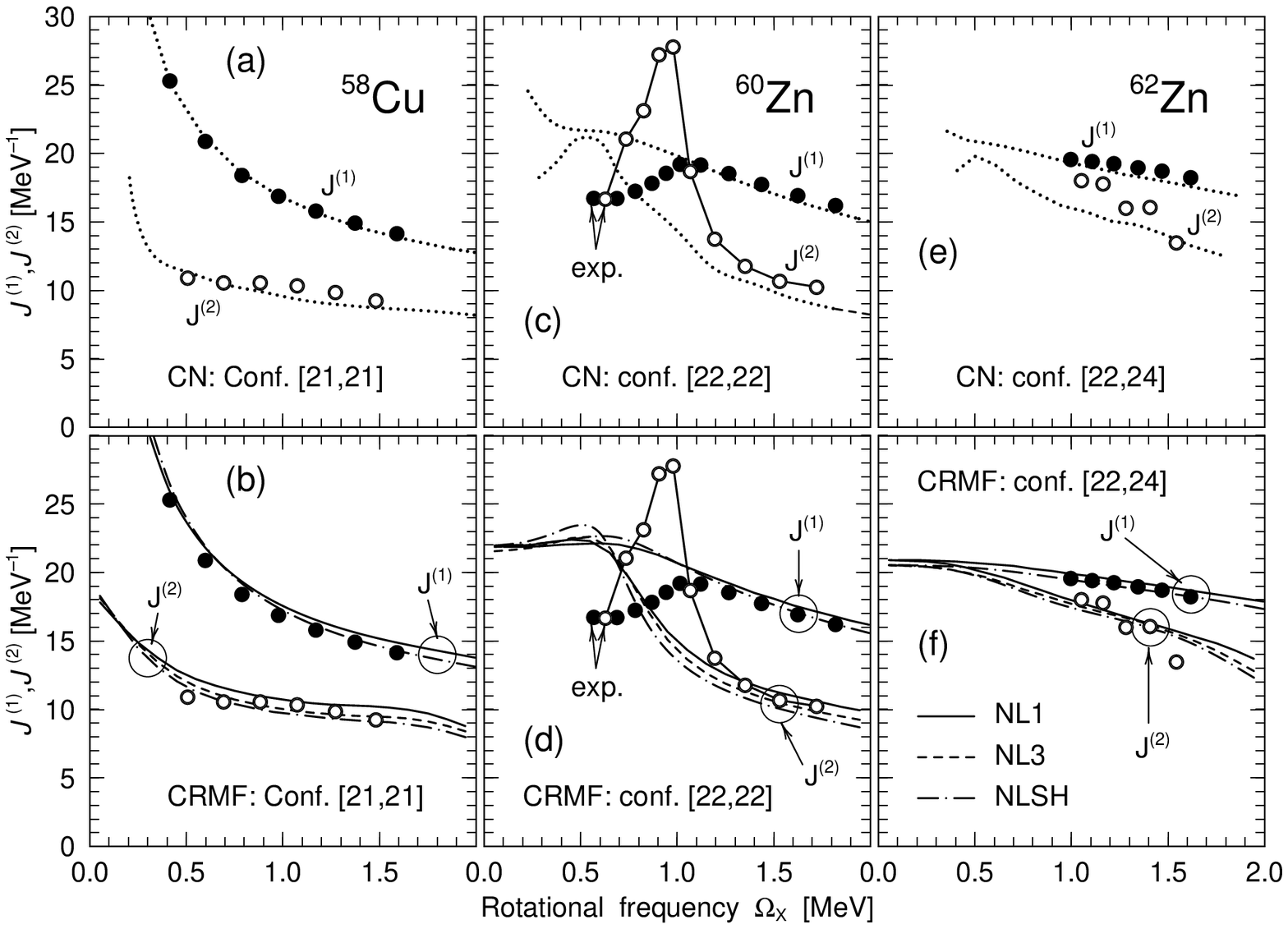}
\caption{Kinematic $J^{(1)}$ (unlinked solid circles) 
and dynamic $J^{(2)}$ (open circles) moments of inertia 
of observed bands versus the ones of assigned calculated 
configurations. The notation of lines is given in the
figure.}
\label{fig-j2j1}
\end{figure}

According to the CRMF and the CN approaches, the doubly-magic 
SD band in $^{60}$Zn has a $[22,22]$ structure in the notation
defined above. In this configuration, all single-particle levels 
below the $Z=N=30$ SD shell gaps are occupied (see Fig.
\ref{routh-crmf}). At the spins of interest, this band is well 
separated from excited SD configurations (see Fig. 3 in 
Ref. \cite{Zn60SD}). The experimental observables 
($J^{(1)}$, $J^{(2)}$ at $\Omega_x \geq 1.1$ MeV and $Q_t$) 
are well described in both approaches (see Figs. \ref{fig-j2j1}c,d
and \ref{fig-eld-qt}). At $\Omega_x \sim 0.95$  MeV, 
the observed band undergoes a paired band crossing, the 
description of which is not addressed in the present 
calculations. Note that this is a $N=Z$ nucleus, so
the proton-neutron pairing correlations could play 
some role at high spin.  
 
\begin{figure}
\epsfxsize 12.5cm
\epsfbox{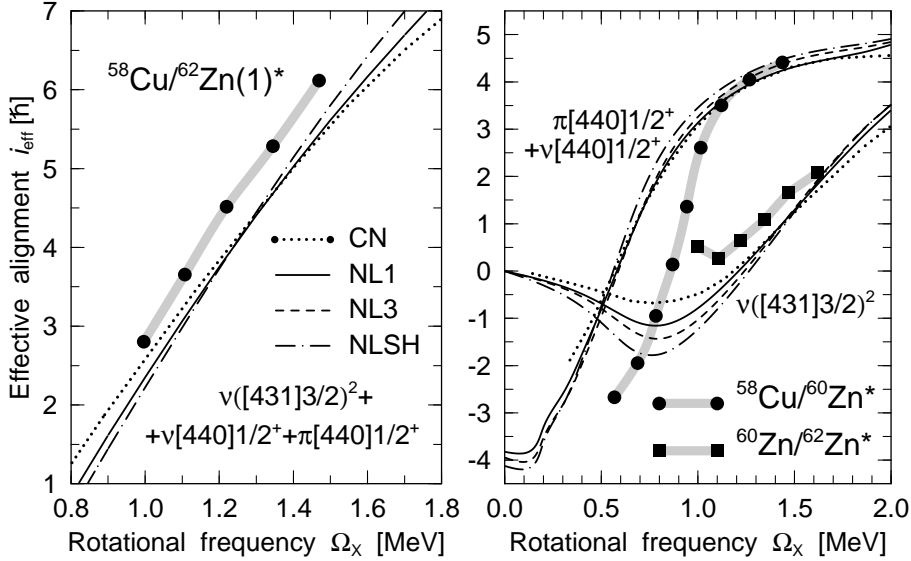}
\caption{Experimental (large solid symbols)  
and calculated (the notation of lines is given in the 
figure) effective alignments. The effective alignment
between bands A and B is defined in Ref. \protect\cite{Rag91} as
$i_{eff}^{A,B}(\Omega_x) = I_{B}(\Omega_x)-I_{A}(\Omega_x)$.
The band A in the lighter nucleus is taken as a reference,
so the effective alignment measures the effect of additional
particles. The experimental effective alignment between bands 
A and B is indicated as ``A/B''. The experimental
$i_{eff}$ values are shown at the transition energies
of the band indicated by an asterisk (*). The compared
configurations differ in the occupation of the orbitals
indicated in the figure.}
\label{fig-align}
\end{figure}

\begin{figure}
\epsfxsize 12.5cm
\epsfbox{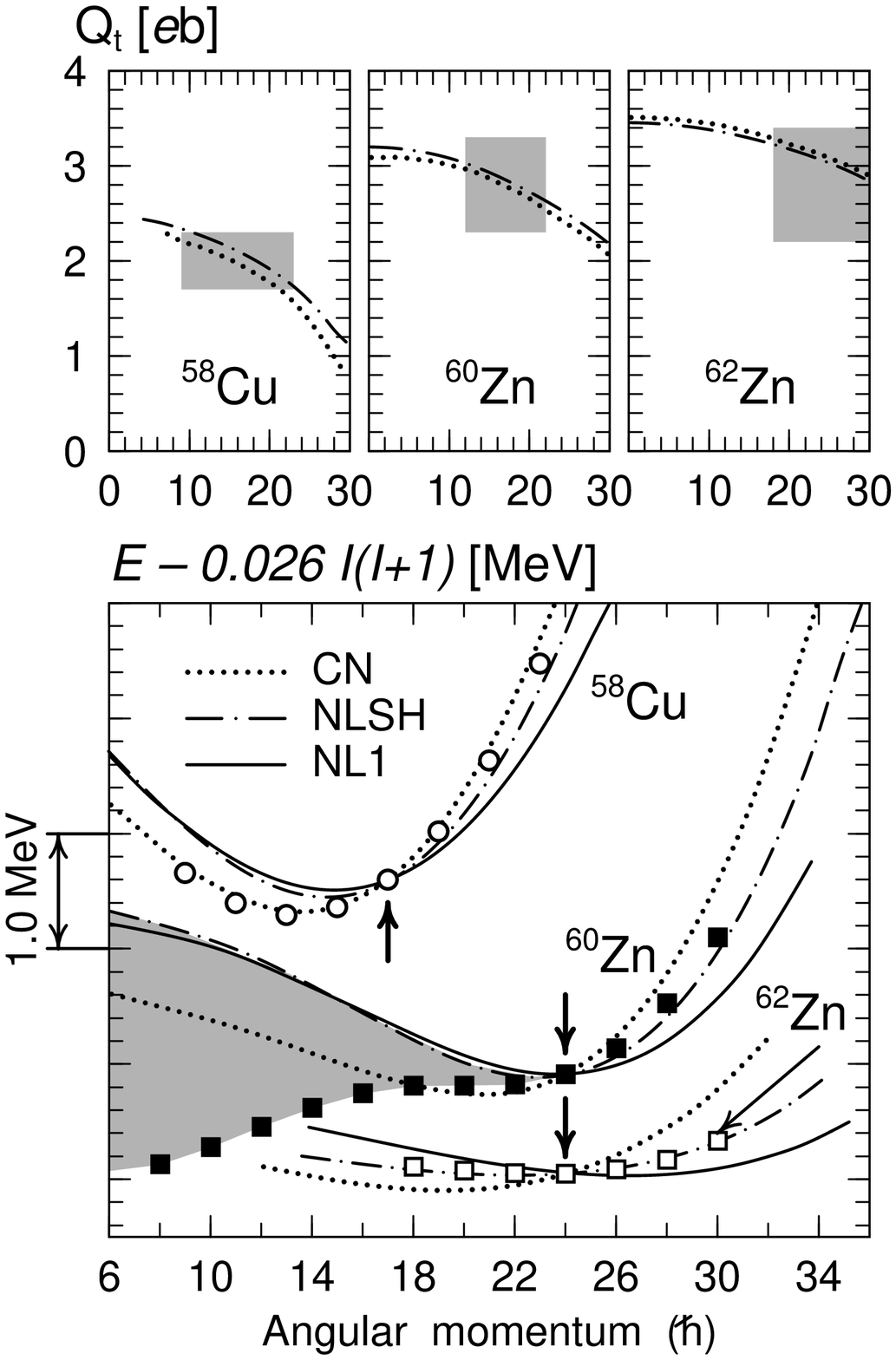}
\caption{Bottom panel: Calculated (lines) and experimental 
(symbols) bands shown 
relative to a rigid rotor reference. The energies of
calculated states indicated by arrows are normalized 
to the corresponding experimental states. The shaded area
is used to indicate the possible size of pairing 
correlations at low spin in the SD band of 
$^{60}$Zn. The results with NL3 are rather close to 
the ones with NLSH, so for simplicity they are not shown.
Top panels: Measured transition quadrupole
moments $Q_t$ (shaded boxes indicate the upper and
lower limits of $Q_t$ and the spin range where they 
have been measured) versus calculated ones. 
Since the $Q_t$ values calculated with 
NL1 and NL3 differ from the ones with NLSH 
only by $\approx2-3\%$, only the results
of calculations with NLSH are shown.}
\label{fig-eld-qt}
\end{figure}

  Contrary to $^{60}$Zn, the bands in $^{58}$Cu and $^{62}$Zn 
are not linked to the low-spin level scheme and thus their 
parities and spins are not known experimentally. Considering 
the available single-particle orbitals in the vicinity of the 
$Z=N=30$ SD shell gaps (see Fig.\ \ref{routh-crmf}) and 
comparing calculated effective alignments $i_{eff}$ 
with observed ones in the $^{58}$Cu/$^{60}$Zn, $^{60}$Zn/$^{62}$Zn, 
$^{58}$Cu/$^{62}$Zn pairs (Fig. \ref{fig-align}), it was found 
that the physical observables ($J^{(1)}$, $J^{(2)}$ and 
$i_{eff}$) are best described when the configurations 
$[21,21]$ and $[22,24]$ are assigned to the bands in 
$^{58}$Cu and $^{62}$Zn, respectively.  The configuration 
$[21,21]$ in $^{58}$Cu is calculated to be energetically 
favored over a considerable spin range in both 
approaches and also in the cranked Hartree-Fock 
approach with Skyrme forces \cite{Cu58}. 

The experimental $J^{(1)}$ and $J^{(2)}$ moments of inertia of 
the $^{62}$Zn band are somewhat better described in CRMF 
theory than in the CN model (see Figs.\ \ref{fig-j2j1}e,f). 
The fact that the last experimental point in $J^{(2)}$ 
is overestimated in the CRMF calculations
is possibly due to an interaction between the
occupied $[431]3/2^+$ and unoccupied 
$[431]1/2^+$ orbitals (see bottom panel of 
Fig.\ \ref{routh-crmf}). One should note, however,
that the configuration $[22,24]$ in $^{62}$Zn is not 
calculated as the lowest SD configuration. In the spin range 
of interest, its energy above the lowest SD solution
is $\approx 1-1.5$ MeV in CRMF theory (see for example Ref. \cite{MM.98})
and $\approx 0.5-1.0$ MeV in the CN model. 
In Ref. \cite{Zn62SD}, the $^{62}$Zn configurations 
$[22,22]$ and $[22,23]$, which are calculated
lowest in energy, were considered instead.
However, especially when compared with the
band in $^{60}$Zn, it becomes evident that these
configurations do not provide a satisfactory description 
of the observed properties of SD band in $^{62}$Zn. The 
configuration with only one neutron hole in the $f_{7/2}$ 
orbital can be excluded for similar reasons but also because
its signature partner is calculated degenerate in
energy contrary to experiment where no  
signature partner band has been observed so far.
 
The calculated high energy of the configuration
assigned to the observed band in $^{62}$Zn 
suggests that the parameterizations used might not be optimal 
with respect of description of single-particle energies 
in the vicinity of the SD shell gaps. Note, however, that 
the CN model and CRMF theory with three different
forces indicate the same group of orbitals in 
the vicinity of the $N=Z=30$ SD shell gap (see
top panel of Fig.\ \ref{routh-crmf}). In both 
approaches, the $N=Z=30$ SD shell gap is primarily 
defined by the energy splitting between the $[440]1/2$ 
and $[431]3/2$ orbitals originating from the intruder 
$g_{9/2}$ subshell. Thus the lowering of the 
$g_{9/2}$ subshell by $\sim 0.5-1$ MeV will almost 
not affect the size of the $N=Z=30$ SD shell gap 
but will bring the conf. $[22,24]$ in $^{62}$Zn
closer to the yrast line. Note that this will
also make the $N=38$ shell gap seen in Fig. \ref{routh-crmf}
smaller due to the lowering of the $[422]5/2$ orbital. The 
observation of other SD structures in $^{62}$Zn 
and neighboring nuclei will be essential to 
establish the ordering of single-particle levels 
around the $N=30$ SD shell gap and to determine the
accuracy with which existing theories describe the 
alignment properties of single-particle orbitals.
This should help to clarify how the models should be 
further improved to give an even better description 
of the variety of rotational structures observed in this 
mass region. 
 
 As inferred from the analysis of effective alignments in 
Fig. \ref{fig-align}, with the present configuration 
assignments, the lowest transition in the highly-deformed band of 
$^{58}$Cu with the transition energy of 830 keV corresponds 
to a spin change of $11^+ \rightarrow 9^+$ and that 
the lowest transition in the SD band of $^{62}$Zn 
with the transition energy of 1993 keV corresponds 
to a spin change of $20^+ \rightarrow 18^+$. Thus the bands 
in $^{58}$Cu and $^{62}$Zn are observed up to $23^+$ and $30^+$, 
respectively. The corresponding experimental values of 
$J^{(2)}$ and $J^{(1)}$ (under these spin assignments) 
are reproduced rather well in the calculations 
(see Fig.\ \ref{fig-j2j1}).

 Considering the distribution of particles and holes 
over high- and low-$j$ orbitals at low spin one obtains
the 'maximum' spins of the configurations of the observed 
bands as $I=29^+$ ($^{58}$Cu), $I=36^{+}$ ($^{60}$Zn) 
and $I=40^+$ ($^{62}$Zn). Thus the bands in $^{58}$Cu and $^{60}$Zn 
($^{62}$Zn) are three (five) transitions away from 
the 'maximum' spin. Note however that the states of 'maximum' 
spin are collective due to the interaction between 
high- and low-$j$ orbitals of the $N=3$ shell, i.e.  
these bands do not terminate in the usual sense.
Even so, the properties of these bands are strongly influenced 
by the limited angular momentum content of their single-particle 
configurations. Indeed, several features of these bands are
similar to those of smooth terminating bands observed in the 
$A\sim 110$ mass region \cite{A110} and in $^{62,64}$Zn 
\cite{Zn62bt,Zn64bt}. Such features are the smooth drop 
of the dynamic moment of inertia $J^{(2)}$ with increasing 
rotational frequency to values much lower than the 
kinematic moment of inertia $J^{(1)}$. Furthermore, 
a gradual drop of collectivity (i.e. a drop of transition
quadrupole moment $Q_{t}$) is predicted with increasing spin
for both kinds of bands, something which at present has been 
experimentally confirmed \cite{Zn62bt,Wad98} only for smooth 
terminating bands. Indeed, in the $A\sim 60$ region, one 
can see the gradual transition from the smooth terminating 
bands in $^{62,64}$Zn over the highly-deformed band in 
$^{58}$Cu to the SD bands in  $^{60,62}$Zn. Calculations 
for a number of configurations in neighboring nuclei and in 
$^{68}$Zn (see Ref.\ \cite{Zn68} for this nucleus) show that 
the above mentioned features are common for the SD and 
highly-deformed bands in the $A\sim 60-70$ 
mass region. Thus a rigid-rotor assumption ($J^{(1)} \approx J^{(2)})$
sometimes used in the analysis of SD bands is not valid
in this mass region. Indeed, in line with previous 
studies \cite{Rag87}, we can conclude that it is not so much 
the deformation (at $I=0$) of
a band which determines if it is rigid-rotor like or not
but rather how far away the band is from its  'maximum' 
spin value. Therefore, it is in general much more difficult 
to find rigid-rotor like rotational bands in light nuclei 
because the 'maximum' spin within the yrast 
and near-yrast configurations is generally much lower
than in heavier nuclei.

 One should note the important role of the first two
$f_{7/2}$ holes (in [303]7/2 orbitals at prolate shape)
in the stabilization of high- and superdeformation 
for the nuclei around $Z=N=30$. Their influence is 
twofold. First, they significantly contribute to 
the quadrupole moment (see Ref.\ \cite{Dob}). In this 
respect the highly-deformed and SD bands in the 
$A\sim 60$ mass region are similar to the ones in 
the $A\sim 135$ mass region, where the proton 
$g_{9/2}$ holes play an important role in 
stabilization of superdeformation (see 
Ref. \cite{A130}). Second, the contribution 
to the 'maximum' spin of these $f_{7/2}$
holes is comparable with the contribution 
from the $g_{9/2}$ particles. For example, 
full alignment of two highest $f_{7/2}$ holes 
gives $6\hbar$, while full alignment of two 
lowest $g_{9/2}$ particle gives $8\hbar$ in angular momentum. 

 The SD and highly-deformed bands in the $A\sim 60$  mass 
region are characterized by very large transition energies 
reaching 3.2 MeV or more at the top of all three bands 
studied here. For these bands, the excitation energies drawn 
relative to rigid rotor reference appears to provide the best 
measure of how well the theory describes the 
response of the nuclei to rotation as illustrated in the 
bottom panel of Fig.\ \ref{fig-eld-qt}. Note, however, that since 
the pairing correlations are neglected in the calculations
the comparison between theory and experiment should be made 
not with respect to the ground state but with respect to
some high spin state. Indeed, in this kind of plot the 
difference between different approaches and different 
parameterizations of the RMF theory is more clearly seen 
compared with the plot of dynamic and kinematic moments of
inertia (see Fig.\ \ref{fig-j2j1}). Comparing different
results one can conclude that the best description of 
excitation energies within configurations
at high spin is obtained within the 
CRMF theory with the NLSH and NL3 forces for $^{60,62}$Zn 
while the band in $^{58}$Cu is somewhat better described
in the CN model. Concerning relative energies of different 
configurations (not shown in Fig. \ref{fig-eld-qt}), none of 
the approaches give the configuration assigned to the SD band
in $^{62}$Zn as yrast with somewhat larger discrepancies
in the CRMF than in the CN approach. 

   In conclusion, the cranked relativistic mean field 
theory and the configuration-dependent shell-correction
approach with the cranked Nilsson potential have been
used for a study of super- and highly-deformed rotational 
bands in the $A\sim 60$ mass region. The experimental 
observables like the dynamic moment of inertia $J^{(2)}$ 
and the transition quadrupole moments $Q_t$ are well 
described by both approaches. Using the fact that 
the yrast SD band in $^{60}$Zn is linked to the 
low-spin level scheme it was shown that by means 
of the effective alignment approach it is possible to
establish absolute spin values for the unlinked 
highly-deformed and superdeformed bands in 
neighboring nuclei. Thus for the first time it becomes 
possible to compare theory with experiment 
in a direct way also for spin-dependent physical 
observables like kinematic moment of inertia, $J^{(1)}$, 
and excitation energies as a function of spin, $E(I)$, 
for superdeformed bands in the unpaired regime. It was 
found that a much lower value of the dynamic $J^{(2)}$ 
than the kinematic $J^{(1)}$  moment of inertia 
at high spin is rather general feature of SD and 
highly-deformed bands in this mass region.

   A.V.A. acknowledges support from the Alexander von 
Humboldt Foundation. This work is also supported in part 
by the Bundesministerium f{\"u}r Bildung und Forschung 
under the project 06 TM 875 and by the Swedish Natural 
Science Research Council.

\end{document}